\documentclass[twoside]{article}
\usepackage{wrapfig}
\usepackage{graphicx}
\usepackage{amssymb}
\usepackage{cmpj2e}

\title[Nonequilibrium Perturbative Formalism 
]{ Nonequilibrium Perturbative Formalism  
 and Spectral Function for the Anderson Model}
  \author[Mami Hamasaki ]{Mami Hamasaki}
  \address{Department of Physics, Kyoto University, Kyoto 606-8502, Japan}
\begin{document}

\maketitle

\begin{abstract}
The present work is based on the nonequilibrium perturbative 
formalism. There the self-energies are derived up to the forth-order. 
In consequence, it proves that the nonequilibrium 
( real-time ) perturbative expansion can be connected with the Matsubara 
imaginary-time perturbative expansion for equilibrium.  
As the numerical results, the Kondo resonance still disappears 
for bias voltage exceeding the Kondo temperatures, as observed 
in experiments of two terminal systems. 


\pacs 71.15.-m, 05.70.Ln
\end{abstract}

\section{Introduction }
\subsection {Nonequilibrium Perturbative Formalism }
The basic idea on the nonequilibrium perturbation theory 
grounded on the time-contour which starts and ends at  
$t=-\infty$ via $t=\infty$ has been proposed by Schwinger.~\cite{1}  
After that, the frame of the nonequilibrium perturbation theory has 
been built up using the nonequilibrium Green's  functions 
given after the time-contour by Keldysh.~\cite{2}  
The perturbative equation is expressed in matrix form:
\begin{eqnarray}
{\bf G}&=&  \ \  {\bf g} \ \  +  \ \ {\bf g}
 \ \ {\bf{\Sigma}} \ \ {\bf G},  
\end{eqnarray}
where 
\begin{eqnarray}
{\bf G}=\left[ \begin{array}
{ll}
G^{--} & G^< \\
 G^> & G^{++} \\
\end{array} 
\right],  \ \
{\bf {\Sigma}}=\left[ \begin{array}
{ll}{\Sigma}^{--} & {\Sigma}^< \\
{\Sigma}^> & {\Sigma}^{++} \\
\end{array} 
    \right].  \nonumber  
\end{eqnarray}
The nonequilibrium Green's functions are given 
 in the Heisenberg representation by 
\begin{eqnarray}
& & G^{--}(t_1,t_2)\,{\equiv}\,-i{\langle}
{\rm T}\hat{d}(t_1)
 \hat{d}^{\dag}(t_2){\rangle},  \\
& &   G^{++}(t_1,t_2)\,{\equiv}\,
-i{\langle}{\rm \tilde{T}}\hat{d}(t_1)
 \hat{d}^{\dag}(t_2){\rangle},   \\
& &    G^> (t_1,t_2) \,{\equiv}\,
-i{\langle}\hat{d}(t_1)
 \hat{d}^{\dag}(t_2){\rangle},  \\
& &   G^< (t_1,t_2)\,{\equiv}\,i
{\langle}\hat{d}^{\dag}(t_2) \hat{d}(t_1){\rangle}.   
\end{eqnarray}
Here, the time ordering operator ${\rm T}$ arranges in 
chronological order and ${\rm \tilde{T}}$ is the anti time 
ordering operator which arranges in the reverse of 
chronological order. The angular brackets 
denote  thermal average in nonequilibrium. 

 The present work is undertaken on the basis of  
the nonequilibrium perturbative formalism, Eq. ( 1 ). 
The retarded and advanced self-energies up to the fourth-order 
are formulated. 
Then it is confirmed that the nonequilibrium perturbative expansion 
can be connected with the Matsubara imaginary-time perturbative 
expansion for equilibrium.  

\subsection {The Kondo effect}
The Kondo effect~\cite{3} was discovered forty years ago
 and after that, the Kondo physics has been  clarified 
from Landau's Fermi liquid theory~\cite{4}, 
the renormalization group~\cite{5}, scaling~\cite{6}, 
etc.. Besides, generalized Kondo problem,  
that of more than one channel or one impurity has been 
investigated.~\cite{7,8} 
Then, the Kondo effect in electron transport through
 a quantum dot has been predicted theoretically
 at the end of 1980s~\cite{9} 
and after a decade, this phenomenon has been observed.~\cite{10} 
The Kondo effect has been studied theoretically using the 
Anderson model and the predictions 
 have been confirmed experimentally. 
In the Kondo regime, the conductance has been observed to reach 
the unitarity limit and the Kondo temperatures estimated 
 from observation~\cite{11} are in excellent  agreement 
with the expression  derived using the Anderson model.~\cite{12} 
Furthermore, the Kondo effect in a quantum dot has been studied 
 for nonequilibrium system where the bias voltage is applied.~\cite{13} 
The Yamada-Yosida theory,~\cite{14} perturbation theory for equilibrium 
based on  the Fermi liquid theory~\cite{4} has been extended to nonequilibrium 
system and it has been shown that for bias voltage higher than 
the Kondo temperatures, the Kondo resonance 
disappears in the spectral function with the second-order 
self-energy of the Anderson model.~\cite{15}
This is in good agreement with the experiments of two terminal systems 
 that it has been observed that the Kondo effect is suppressed when 
source-drain bias voltage is comparable to or exceeds 
the Kondo temperatures.~\cite{16,17} 

In the present work, using self-energies derived up to the 
fourth-order, the behavior of Kondo resonance is investigated for 
nonequilibrium state caused by bias voltage. 
The numerical results are still that the Kondo resonance is broken, 
as supported by two-terminal experiments. 

\section{ Nonequilibrium  Perturbation Theory }
\subsection{ Formalism }

A thermal average can be obtained 
on the basis of the nonequilibrium perturbation 
 theory.~\cite{1,2,18,19,20,21,22} 
It is assumed that we can know only the state at $t=-{\infty}$, 
that is, initially at $t=-{\infty}$, the system is 
equilibrium and/or noninteracting state. 
The perturbation is turned on at $t=-{\infty}$     
and introduced adiabatically and then, brought wholly 
into the system at $t=0$; around $t=0$, the system is 
regarded as stationary nonequilibrium and/or interacting state. 
After that, the perturbation  
is taken away adiabatically and vanishes at $t={\infty}$. 
When the time evolution of the state is 
irreversible, then, the state at $t={\infty}$ cannot be 
well-defined.  
 The time evolution is therefore, performed 
along the real-time contour which starts and ends 
at $t=-\infty$, as illustrated in Fig. 1.

\begin{figure}[htbp]
\begin{center}
\includegraphics[width=10.0cm,clip=true]{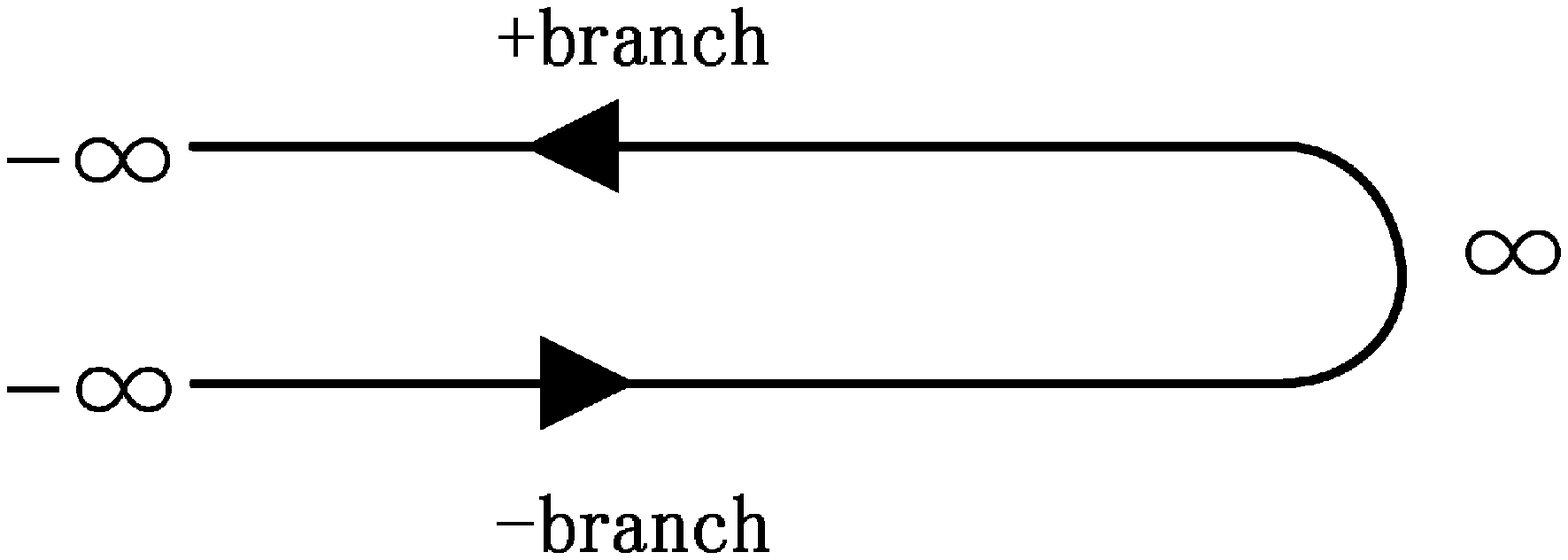}
\end{center}
\caption{The time-contour which starts and ends 
at $t=-\infty$.}
\end{figure}

  S matrix is defined by 
\begin{eqnarray}
{\cal S}(t,t_{\rm 0})&=&1+\sum_{n=1}^{\infty}
{ \frac {1}{n!} }\left({ \frac{-i}{\hbar} }\right)^n
\int_{t_{\rm 0}}^t dt_1 {\ldots} 
\int_{t_{\rm 0}}^t dt_n 
{\rm T}\left[{\tilde{\cal {H}}_{\rm I}}(t_1){\ldots}
{\tilde{\cal {H}}_{\rm I}}(t_n)\right] \nonumber   \\
&=&{\rm T}\left[ {\exp} \left\{ { \frac{-i}{\hbar} }\int_{t_{\rm 0}}^t 
dt^{'}{\tilde{\cal {H}}_{\rm I}}(t^{'}) \right\} \right],   \\ 
{\cal S}(t,t_{\rm 0})^{\dag}&=&{\cal S}(t_{\rm 0},t)  
={\tilde{\rm T}}\left[ {\exp} \left\{ { \frac{i}{\hbar} }
\int_{t_{\rm 0}}^t dt^{'}{\tilde{\cal {H}}_{\rm I}}(t^{'}) \right\} \right].  
\end{eqnarray}
Here ${\tilde{\cal {H}}_{\rm I}}$ is perturbation term 
in interaction representation. 

For thermal equilibrium, the statistical operator ( density matrix ) is 
written in Gibbs form for the grand canonical ensemble by 
\begin{eqnarray}
{\varrho}_{G}={\frac{e^{-{\beta}({\cal H}-{\mu}N)}}
{{\rm Tr}e^{-{\beta}({\cal H}-{\mu}N)}}}=e^{{\beta}({\Omega}-{\cal H}+{\mu}N)}.
\end{eqnarray}
Equation ( 8 ) is not valid exactly for nonequilibrium.
We have no specific limitations upon the statistical operator. 
The statistical operator can generally be expressed 
in Schr{\"o}dinger representation by~\cite{20,22} 
\begin{eqnarray}
{\varrho}_S(t)=\sum_{m}|m_S(t)>P_m<m_S(t)|. 
\end{eqnarray}
Here, $P_m$ is probability that the system is in  
 state $m$ and $|m_S(t)>$ is the state in Schr{\"o}dinger 
representation. ${\varrho}_S$ satisfies the Liouville 
equation by 
\begin{eqnarray}
i{\hbar}{\frac{ {\partial}{\varrho}_S }{ {\partial} t }}
=[{\cal {H}},{\varrho}_S]. 
\end{eqnarray}
The statistical operator in the interaction representation is given by  
${\tilde{\varrho}(t)}=e^{i{\cal {H}_{\rm 0}}t/ {\hbar}} 
{\varrho}_S(t)e^{-i{\cal {H}_{\rm 0}}t/{\hbar}}$ and satisfies 
\begin{eqnarray}
i{\hbar}{\frac{ {\partial}{\tilde{\varrho}} }{ {\partial} t }}
&=&[{\tilde{\cal {H}}}_{\rm I},{\tilde{\varrho}}]. 
\end{eqnarray}
As a matter of course, ${\varrho}_S(0)={\varrho}(0)={\tilde{\varrho}(0)}. $  
Here ${{\varrho}(t)}$ is in the Heisenberg representation. 
The time evolution is expressed using S matrix by 
\begin{eqnarray}
{\tilde{\varrho}}(t)=S(t,t_{\rm 0}){\tilde{\varrho}}(t_0)S(t_{\rm 0},t).
\end{eqnarray}

 The thermal average in the Heisenberg representation 
 at $t=0$ can be obtained, for example by~\cite{19,22} 
\begin{eqnarray}
 & & {\langle}{\rm T} A(t)B(t^{'}){\rangle}  \nonumber   \\
&{\equiv}&{\rm Tr} [ {\varrho}(0){\rm T}A(t)B(t^{'})]  
\nonumber   \\
&=&{\rm Tr} [ {\tilde{\varrho}}(-\infty)
{\cal S}(-\infty,0){\rm T}A(t)B(t^{'}){\cal S}(0,-\infty)]
\nonumber   \\
&=&{\rm Tr} [ {\tilde{\varrho}}(-\infty)
{\cal S}(-\infty,\infty)\{{\rm T}{\cal S}(\infty,-\infty)
{\tilde{A}(t^-)}{\tilde{B}(t^{'-})}\}]
\nonumber  \\ 
&=&\sum_{n=1}^{\infty}\sum_{m=1}^{\infty}
{ \frac {1}{n!} }{ \frac {1}{m!} }
\left({ \frac{i}{\hbar} }\right)^n
\left({ \frac{-i}{\hbar} }\right)^m
\int_{-\infty}^{\infty} dt_1 {\ldots} 
\int_{-\infty}^{\infty} dt_n 
\int_{-\infty}^{\infty} dt_1^{'} {\ldots} 
\int_{-\infty}^{\infty} dt_m^{'} 
\nonumber   \\ 
& &{\times}{\langle}\left\{ {\tilde{\rm T}}
{\tilde{\cal {H}}}_{\rm I}(t_1^{+})
{\ldots}{\tilde{\cal {H}}}_{\rm I}(t_n^{+}) \right\}\left\{ {\rm T}
 {\tilde{\cal {H}}}_{\rm I}(t_1^{'-})
{\ldots}{\tilde{\cal {H}}}_{\rm I}(t_m^{'-})
{\tilde{A}(t^{-})}{\tilde{B}(t^{'-})} \right\}{\rangle}_{av}, 
\nonumber
\end{eqnarray}
where  
${\langle}{\ldots}{\rangle}_{av}
$$={\rm Tr}[{\tilde{\varrho}}(-\infty){\ldots}].$
Then, the thermal average is derived 
by following the ordinary procedure via the Wick's theorem. 

\subsection{ Relation of Self-Energy }
After the perturbative expansion is executed,  
the retarded and advanced self-energies are formulated.
According to the definition, the retarded and advanced 
Green's functions  are given by  
\begin{eqnarray}
 & & G^r (t_1,t_2)\,{\equiv}\,-i{\theta}(t_1-t_2)
 {\langle}\{\hat{d}(t_1),
 \hat{d}^{\dag}(t_2)\}{\rangle},  \\
& &  G^a (t_1,t_2)\,{\equiv}\,i{\theta}(t_2-t_1)
 {\langle}\{\hat{d}(t_1),
 \hat{d}^{\dag}(t_2)\}{\rangle}.  
\end{eqnarray}
Here, the curly brackets denote anticommutator.
The Dyson's equations for the retarded and advanced 
Green's functions are given by 
\begin{eqnarray}
G^r&=&  \ \   g^r \ \  +  \ \  g^r
 \ \ {\Sigma}^r \ \  G^r,  \\
G^a&=&  \ \   g^a \ \  +  \ \  g^a
 \ \ {\Sigma}^a \ \  G^a.  
\end{eqnarray} 
As the necessity to Eqs. ( 15 ) and ( 16 ),  
the self-energies ${\Sigma}^r$ and ${\Sigma}^a$ 
are also required to be retarded and advanced functions in time, respectively.  
 In accordance with the ordinary procedure of nonequilibrium perturbative 
formalism,~\cite{2,19} there 
\begin{eqnarray}
{\bf L}=[{\bf L^{\dag}}]^{-1}={\frac {1}{\sqrt{2}}}\left[ \begin{array}
{ll}
1 & -1 \\
1 & 1 \\
\end{array} 
\right], \nonumber  
\end{eqnarray}
and using this, then, 
\begin{eqnarray}
{\bf {\Sigma}}=\left[ \begin{array}
{ll}{\Sigma}^{--} & {\Sigma}^< \\
{\Sigma}^> & {\Sigma}^{++} \\
\end{array} 
    \right]  \ \ {\longrightarrow} \ \ 
{\bf L}{\bf {\Sigma}}{\bf L^{\dag}}=\left[ \begin{array}
{ll}{\Omega} & {\Sigma}^r \\
{\Sigma}^a & 0 \\
\end{array} 
    \right].  \nonumber  
\end{eqnarray}
The relationship for self-energies ought to be obtained here 
by comparison of Eq. ( 1 ) with Eqs. ( 15 ) and ( 16 ):  
\begin{eqnarray}
{\Sigma}^{r}(t)&=&{\Sigma}^{--}(t)+{\Sigma}^{<}(t)
=-{\Sigma}^{++}(t)-{\Sigma}^{>}(t), \\
{\Sigma}^{a}(t)&=&{\Sigma}^{--}(t)+{\Sigma}^{>}(t)
=-{\Sigma}^{++}(t)-{\Sigma}^{<}(t).
\end{eqnarray}

\section{ Expressions of Self-Energy for Anderson model }
\subsection{ Anderson Model }

We consider equilibrium and nonequilibrium stationary states.  
Nonequilibrium state is caused by finite bias voltage, 
that is, the difference of chemical potentials;    
after bias voltage was turned on, long time has passed enough 
to reach stationary states. 
Since the states are stationary, Hamiltonian has no 
time dependence. The system is  described by 
the Anderson model connected to leads. 
The impurity with on-site energy $E_0$ and the Coulomb interaction $U$ 
is connected to the left and right leads   
by the mixing matrix elements, $v_L$ and $v_R$.  
 The Anderson Hamiltonian is given by 
\begin{eqnarray}
{\cal H}=& & E_0 \sum_{\sigma}\hat{n}_{d{\sigma}}
+ {\mu}_L \sum_{\sigma}\hat{n}_{L{\sigma}}
+ {\mu}_R \sum_{\sigma}\hat{n}_{R{\sigma}}
+U \bigl(\hat{n}_{d\uparrow}
-{\langle}\hat{n}_{d\uparrow}{\rangle}\bigr)
\bigl(\hat{n}_{d\downarrow}-{\langle}\hat{n}_{d\downarrow}{\rangle}\bigr)
\nonumber  \\
  & & -\sum_{{\sigma}}v_{L}\bigl(
  \hat{d}_{{\sigma}}^{\dag}\hat{c}_{L{\sigma}}+{\rm H.c.}\bigr) 
-\sum_{{\sigma}}v_{R}\bigl(\hat{d}_{{\sigma}}^{\dag}\hat{c}_{R{\sigma}}
+{\rm H.c.}\bigr). 
\end{eqnarray}
$\hat{d}^{\dag}$ ($\hat{d}$) is creation (annihilation) operator 
 for electron on the impurity, and 
$\hat{c}_{L}^{\dag}$ and $\hat{c}_{R}^{\dag}$ 
($\hat{c}_{L}$ and $\hat{c}_{R}$) are creation (annihilation) operators 
in the left and  right leads, respectively. 
${\sigma}$ is index for spin.  
The chemical potentials in the isolated left and right leads  
are ${\mu}_{L}$ and ${\mu}_{R}$, respectively. 
  The applied voltage is, therefore 
 defined by $eV\, {\equiv}\, {\mu}_L-{\mu}_R$.

We consider that the band-width of left and right leads 
is large infinitely, so that the coupling functions, 
 ${\Gamma}_L$ and ${\Gamma}_R$ can be taken to be independent of energy, $E$.
On-site energy $E_0$ is set canceling with 
the Hartree term, {\it i.e.} the first-order contribution to self-energy for  
electron correlation: 
${\Sigma}^{r(1)}_{{\sigma}}(E)$$={\Sigma}^{a(1)}_{{\sigma}}(E)$
$=U {\langle}n_{-{\sigma}}{\rangle}$.  

Accordingly, the Fourier components of the noninteracting 
 ( unperturbed ) Green's functions reduce to  
\begin{eqnarray}
 & & g^r(E)={\frac{1}{E+i{\Gamma},}} \\
 & & g^a(E)={\frac{1}{E-i{\Gamma},}} 
\end{eqnarray}
where ${\Gamma}=( {\Gamma}_L+{\Gamma}_R )/ 2$. 
Hence, the inverse Fourier components  
 can be written by 
\begin{eqnarray}
g^r(t)&=&-i{\theta}(t)e^{-{\Gamma}t}, \nonumber \\
g^a(t)&=&i{\theta}(-t)e^{{\Gamma}t}.   
\nonumber 
\end{eqnarray}
In addition,  from Eq. ( 1 ), we have    
\begin{eqnarray}
 & & g^<(E)=g^r(E)\, 
\bigl[\, if_L(E){\Gamma}_L+if_R(E){\Gamma}_R \, \bigr]\, g^a(E), \\
 & & g^>(E)=g^r(E)\, 
\bigl[\, i(f_L(E)-1){\Gamma}_L+i(f_R(E)-1){\Gamma}_R \, \bigr]\, g^a(E). 
\end{eqnarray}
$f_L$ and $f_R$ are the Fermi distribution functions 
in the isolated left and right leads, respectively. 
By Eqs. ( 22 ) and ( 23 ), the nonequilibrium state is  
introduced as the superposition of the left and right leads.  
Then, the effective Fermi distribution function can be 
expressed by~\cite{15}
\begin{eqnarray}
f_{\rm eff}(E)={\frac {f_L(E){\Gamma}_L+f_R(E){\Gamma}_R}
{{\Gamma}_L+{\Gamma}_R}}.
\end{eqnarray}

\subsection{ Self-Energy  }
The retarded and advanced self-energies are derived up to the 
forth-order. 
Equations ( 17 ) and ( 18 ) are divided into 
retarded and advanced terms in time: 
\begin{eqnarray}
{\Sigma}^{r}(t)&=&[{\Sigma}^{--}(t)+{\Sigma}^{<}(t)]
{\theta}(t)+[{\Sigma}^{--}(t)+{\Sigma}^{<}(t)]{\theta}(-t) \nonumber \\
&=&-[{\Sigma}^{++}(t)+{\Sigma}^{>}(t)]{\theta}(t)
-[{\Sigma}^{++}(t)+{\Sigma}^{>}(t)]{\theta}(-t), \nonumber \\
{\Sigma}^{a}(t)&=&[{\Sigma}^{--}(t)+{\Sigma}^{>}(t)]{\theta}(t)
+[{\Sigma}^{--}(t)+{\Sigma}^{>}(t)]{\theta}(-t) \nonumber \\
&=&-[{\Sigma}^{++}(t)+{\Sigma}^{<}(t)]{\theta}(t)
-[{\Sigma}^{++}(t)+{\Sigma}^{<}(t)]{\theta}(-t). \nonumber 
\end{eqnarray}
Then it is found that for the self-energies drawn  
using the Wick's theorem, 
\begin{eqnarray}
{\Sigma}^{--}(t){\theta}(t)=-{\Sigma}^{>}(t){\theta}(t),  & &     
{\Sigma}^{++}(t){\theta}(t)=-{\Sigma}^{<}(t){\theta}(t), \nonumber \\  
{\Sigma}^{--}(t){\theta}(-t)=-{\Sigma}^{<}(t){\theta}(-t),  & & 
{\Sigma}^{++}(t){\theta}(-t)=-{\Sigma}^{>}(t){\theta}(-t). \nonumber 
\end{eqnarray} 
It leads to 
\begin{eqnarray}
& & [{\Sigma}^{--}(t)+{\Sigma}^{<}(t)]{\theta}(-t)= 
-[{\Sigma}^{++}(t)+{\Sigma}^{>}(t)]{\theta}(-t)=0,  \nonumber \\
& & [{\Sigma}^{--}(t)+{\Sigma}^{>}(t)]{\theta}(t)=
-[{\Sigma}^{++}(t)+{\Sigma}^{<}(t)]{\theta}(t)=0. \nonumber 
\end{eqnarray}
As a consequence,  
the retarded and advanced self-energies are obtained 
as retarded and advanced functions of time, respectively: 
\begin{eqnarray}
{\Sigma}^{r}(t)&=&[{\Sigma}^{--}(t)+{\Sigma}^{<}(t)]{\theta}(t)
=-[{\Sigma}^{++}(t)+{\Sigma}^{>}(t)]{\theta}(t), \nonumber \\
{\Sigma}^{a}(t)&=&[{\Sigma}^{--}(t)+{\Sigma}^{>}(t)]{\theta}(-t)
=-[{\Sigma}^{++}(t)+{\Sigma}^{<}(t)]{\theta}(-t) \nonumber.
\end{eqnarray}
In addition, it proves  
\begin{eqnarray}
{\Sigma}^{r}(t)&=&[{\Sigma}^{<}(t)-{\Sigma}^{>}(t)]{\theta}(t), 
\nonumber \\ 
{\Sigma}^{a}(t)&=&[{\Sigma}^{>}(t)-{\Sigma}^{<}(t)]{\theta}(-t). \nonumber
\end{eqnarray}
Hence the following relations still stand:  
${\Sigma}^{r}-{\Sigma}^{a}={\Sigma}^{<}-{\Sigma}^{>}$  
 and furthermore 
\begin{eqnarray}
G^<=(1+G^r{\Sigma}^{r})g^<(1+G^a{\Sigma}^{a})-
G^r{\Sigma}^{<}G^a,  \nonumber \\
G^>=(1+G^r{\Sigma}^{r})g^>(1+G^a{\Sigma}^{a})-
G^r{\Sigma}^{>}G^a. \nonumber    
\end{eqnarray}

As the results, the second-order self-energy  is  written by  
\begin{eqnarray}
{\Sigma}^{r(2)}(E) =
U^2\int^{\infty}_{0}{dt_{1}}    
e^{iEt_{1}}\left[ 
\begin{array}
{llll}
\ \ g^{>}(t_{1})g^>(t_{1}) g^<(-t_{1}) \\ 
- g^<(t_{1})g^<(t_{1}) g^>(-t_{1})   
\end{array} \right] \nonumber \\ 
=U^2\int^{\infty}_{0}{dt_{1}}    
e^{iEt_{1}}\left[ 
\begin{array}
{llll}
\ \ g^{\pm}(t_{1})g^>(t_{1}) g^<(-t_{1}) \\ 
+ g^<(t_{1})g^{\pm}(t_{1}) g^>(-t_{1})   \\
+ g^<(t_{1})g^>(t_{1}) g^{\pm}(-t_{1})
\end{array} \right],  
\end{eqnarray}
\begin{eqnarray}
{\Sigma}^{a(2)}(E) =
U^2\int^{0}_{-\infty}{dt_{1}}    
e^{iEt_{1}}\left[ 
\begin{array}
{llll}
\ \ g^<(t_{1})g^{<}(t_{1}) g^>(-t_{1})   \\
- g^>(t_{1})g^>(t_{1}) g^{<}(-t_{1})
\end{array} \right] \nonumber \\  
=U^2\int^{0}_{-\infty}{dt_{1}}    
e^{iEt_{1}}\left[ 
\begin{array}
{llll}
\ \ g^{\pm}(t_{1})g^>(t_{1}) g^<(-t_{1}) \\ 
+ g^<(t_{1})g^{\pm}(t_{1}) g^>(-t_{1})   \\
+ g^<(t_{1})g^>(t_{1}) g^{\pm}(-t_{1})
\end{array} \right].  
\end{eqnarray}
Here $g^{\pm}(t)=g^r(t)+g^a(t)$, that is, 
$g^{+}(t)=g^r(t)=-i{\theta}(t)e^{-{\Gamma}t}$ 
for $t{\ge}0$  and $g^{-}(t)=g^a(t)=i{\theta}(-t)e^{{\Gamma}t}$ 
for $t<0$. Additionally, $g^<(t)$  and $g^>(t)$ 
 are the inverse Fourier components of Eqs. ( 22 ) and ( 23 ). 
Figure 2 shows the diagram for the second-order self-energy. 
As shown numerically later, the second-order contribution
 coincide with those derived by Hershfield {\it et al.}~\cite{15}. 
In the symmetric equilibrium case,
the asymptotic behavior at low energy is expressed by 
\begin{eqnarray}
 {\Sigma}^{r(2)}(E){\simeq}
-{\Gamma}\left(3- {\frac{{\pi}^2}{4}}\right)
\left(\frac{U}{{\pi}{\Gamma}}\right)^2 \frac{E}{\Gamma}
-i\frac{\Gamma}{2}\left(\frac{U}{{\pi}{\Gamma}}\right)^2
\left(\frac{E}{\Gamma}\right)^2,
\end{eqnarray}
the exact results based on the Bethe ansatz method.~\cite{23,24} 
\begin{figure}[htbp]
\begin{center}
\includegraphics[width=7.0cm]{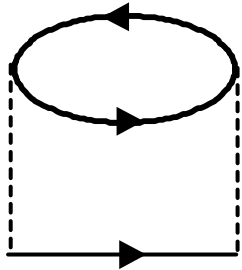}
\end{center}
\caption{The diagram for the second-order self-energy. 
The solid line denotes the noninteracting Green's function 
and the dashed line denotes interaction.
}
\end{figure}

The third-order terms corresponding to the diagram 
 in Fig. 3(a) are expressed by   
\begin{eqnarray}
 {\Sigma}^{r(3)}_{pp}(E)
&=&U^3\int^{\infty}_{0} {dt_{1}}\int^{\infty}_{-\infty}
{dt_{2}}e^{iEt_1}\, 
\left[ 
\begin{array}
{llll}
 g^<(-t_1) g^>(t_1-t_2)
 g^>(t_1-t_2) \nonumber \\ 
 -  g^>(-t_1) g^<(t_1-t_2)
 g^<(t_1-t_2) \nonumber \\ 
\end{array} 
\right] \\
& & {\times} 
\left[ 
\begin{array}
{llll}
 g^{\pm}(t_2) g^>(t_2)  
 + g^<(t_2) g^{\pm}(t_2)  
\end{array} 
\right], 
\end{eqnarray}
\begin{eqnarray}
 {\Sigma}^{a(3)}_{pp}(E)
&=&U^3\int^{0}_{-\infty} {dt_{1}}\int^{\infty}_{-\infty}
{dt_{2}}e^{iEt_1}\, 
\left[ 
\begin{array}
{llll}
g^>(-t_1) g^<(t_1-t_2)
 g^<(t_1-t_2) \nonumber \\
- g^<(-t_1) g^>(t_1-t_2)
 g^>(t_1-t_2)  
\nonumber \\ 
\end{array} 
\right] \\
& & {\times} 
\left[ 
\begin{array}
{llll}
 g^{\pm}(t_2) g^>(t_2)  
 + g^<(t_2) g^{\pm}(t_2)  
\end{array} 
\right]. 
\end{eqnarray}
Figure 3(b) illustrates the diagram for the following terms: 
\begin{eqnarray}
 {\Sigma}^{r(3)}_{ph}(E)
&=&U^3\int^{\infty}_{0} {dt_{1}}\int^{\infty}_{-\infty}
{dt_{2}}e^{iEt_1}\, 
\left[ 
\begin{array}
{llll}
 g^>(t_1) g^>(t_1-t_2)
 g^<(t_2-t_1) \nonumber \\ 
 -  g^<(t_1) g^<(t_1-t_2)
 g^>(t_2-t_1) \nonumber \\ 
\end{array} 
\right] \\
& &{\times} 
\left[ \begin{array}
{llll}
 g^{\pm}(t_2) g^<(-t_2)  
 + g^<(t_2) g^{\pm}(-t_2)  
\end{array} 
\right], 
\end{eqnarray}
\begin{eqnarray}
 {\Sigma}^{a(3)}_{ph}(E)
&=&U^3\int^{0}_{-\infty} {dt_{1}}\int^{\infty}_{-\infty}
{dt_{2}}e^{iEt_1}\, 
\left[ 
\begin{array}
{llll}
 g^<(t_1) g^<(t_1-t_2)
 g^>(t_2-t_1) \nonumber \\ 
- g^>(t_1) g^>(t_1-t_2)
 g^<(t_2-t_1) \nonumber \\ 
\end{array} 
\right] \\
& &{\times} 
\left[ \begin{array}
{llll}
 g^{\pm}(t_2) g^<(-t_2)  
 + g^<(t_2) g^{\pm}(-t_2)  
\end{array} 
\right]. 
\end{eqnarray}
Equations ( 28 )-( 31 ) for equilibrium agree exactly with those   
derived from the  Matsubara imaginary-time perturbative expansion  
for equilibrium and analytical continuity by 
Zlati{\'c} {\it et al.}.~\cite{25}  
As mentioned later, it is numerically confirmed that 
the third-order contribution vanishes for the symmetric Anderson model;  
this is in good agreement with both the results derived from the 
Yamada-Yosida theory~\cite{14,24,26} and obtained 
on the basis of the Bethe ansatz method.~\cite{23}  

\begin{figure}[htbp]
\begin{center}
\includegraphics[width=10.0cm]{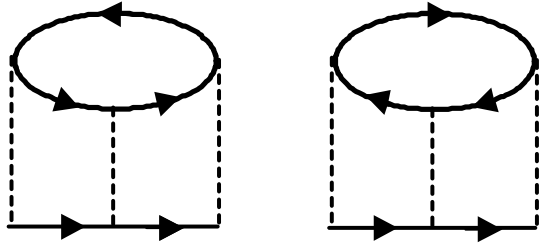}
\end{center}
\caption{The two diagrams for the third-order self-energy. 
Left:(a) and Right:(b)}
\end{figure}
\begin{figure}[htbp]
\begin{center}
\includegraphics[width=7.0cm]{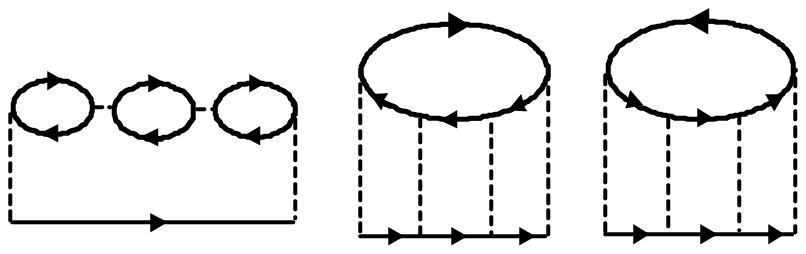}
\includegraphics[width=7.0cm]{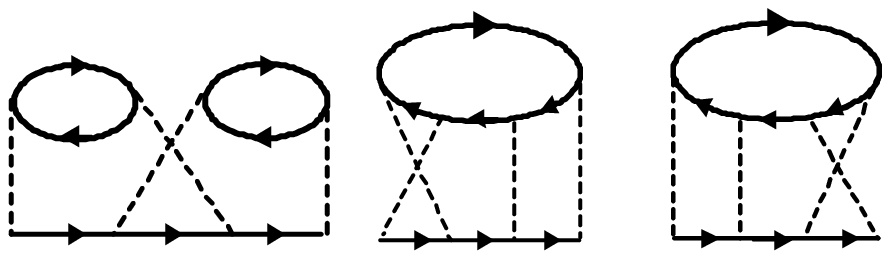}
\includegraphics[width=7.0cm]{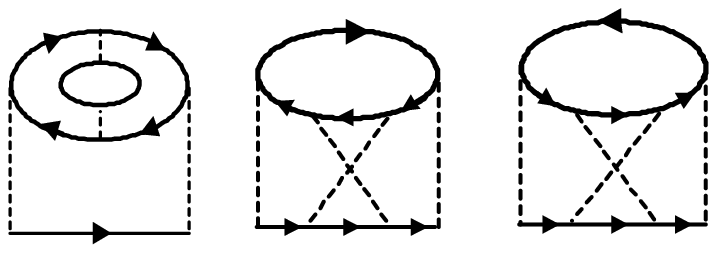}
\includegraphics[width=7.0cm]{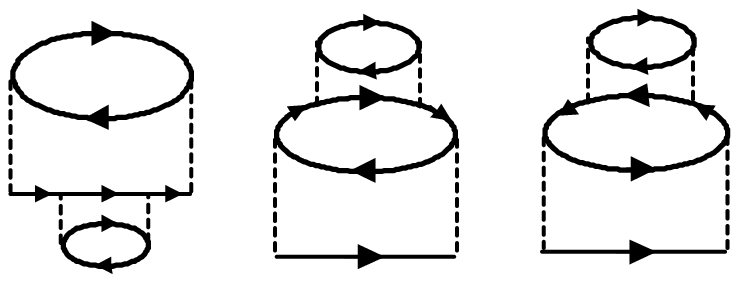}
\caption{The twelve terms for the proper fourth-order self-energy 
 divided into four groups: (a)-(c),  (d)-(f), (g)-(i), 
and (j)-(l). }
\end{center}
\end{figure}

Furthermore, the fourth-order contribution to the 
self-energy is formulated. ( See Appendix. )
The twelve terms for the proper fourth-order self-energy 
 can be divided into four groups, each of which comprises   
 three terms. 
 The four groups correspond to the diagrams 
 denoted in Figs. 4 (a)-(c),  Figs. 4 (d)-(f), Figs. 4 (g)-(i), 
and Figs. 4  (j)-(l), respectively. 
For symmetric Anderson model at equilibrium, 
the asymptotic behavior at low energy  
is approximately in agreement with those based on 
the Bethe ansatz method~\cite{23}:
\begin{eqnarray}
 {\Sigma}^{r(4)}(E){\simeq}
-{\Gamma}\left(105- {\frac{45{\pi}^2}{4}}
+{\frac{{\pi}^4}{16}}\right)
\left(\frac{U}{{\pi}{\Gamma}}\right)^4 \frac{E}{\Gamma}
-i\frac{\Gamma}{2}\left(30-3{\pi}^2\right)\left(\frac{U}
{{\pi}{\Gamma}}\right)^4
\left(\frac{E}{\Gamma}\right)^2.  \nonumber \\
\end{eqnarray}

\section{ Numerical Results and Discussion }
\subsection{ Self-Energy }
The third-order terms, Eqs. ( 28 )-( 31 ) cancel 
under electron-hole symmetry  
 not only at equilibrium but also at nonequilibrium: 
${\Sigma}^{r(3)}_{ph}(E)=-{\Sigma}^{r(3)}_{pp}(E)$  
 and ${\Sigma}^{a(3)}_{ph}(E)=-{\Sigma}^{a(3)}_{pp}(E)$.   
As a consequence, the third-order contribution to self-energy 
vanishes in the symmetric case.  
It agrees with the results of Refs. [14, 24, 26] based on the 
Yamada-Yosida theory that all odd-order contributions 
except the Hartree term vanish for equilibrium
 in the symmetric single-impurity Anderson model;  
probably, it is just the same with nonequilibrium state.
 On the other hand, the third-order terms contribute to 
the asymmetric system where electron-hole symmetry breaks  
and furthermore, the third-order terms for spin-up 
and for spin-down contribute respectively when 
the spin degeneracy is lifted for example, by  magnetic field. 
For the fourth-order contribution,  
three terms which constitute each of four groups 
contribute equivalently under electron-hole symmetry. 
Moreover,  to the asymmetric system, 
the terms brought by the diagrams of Figs. 4(a) and 4(b) 
contribute equivalently  and 
the terms by the diagrams of Figs. 4(j) and 4(k)
make equivalent contribution, and  
the rest, the eight terms contribute respectively. 
Further, the twenty-four terms for spin-up and spin-down 
take effect severally in the presence of magnetic field. 

\begin{figure}[htbp]
\includegraphics[width=8.0cm]{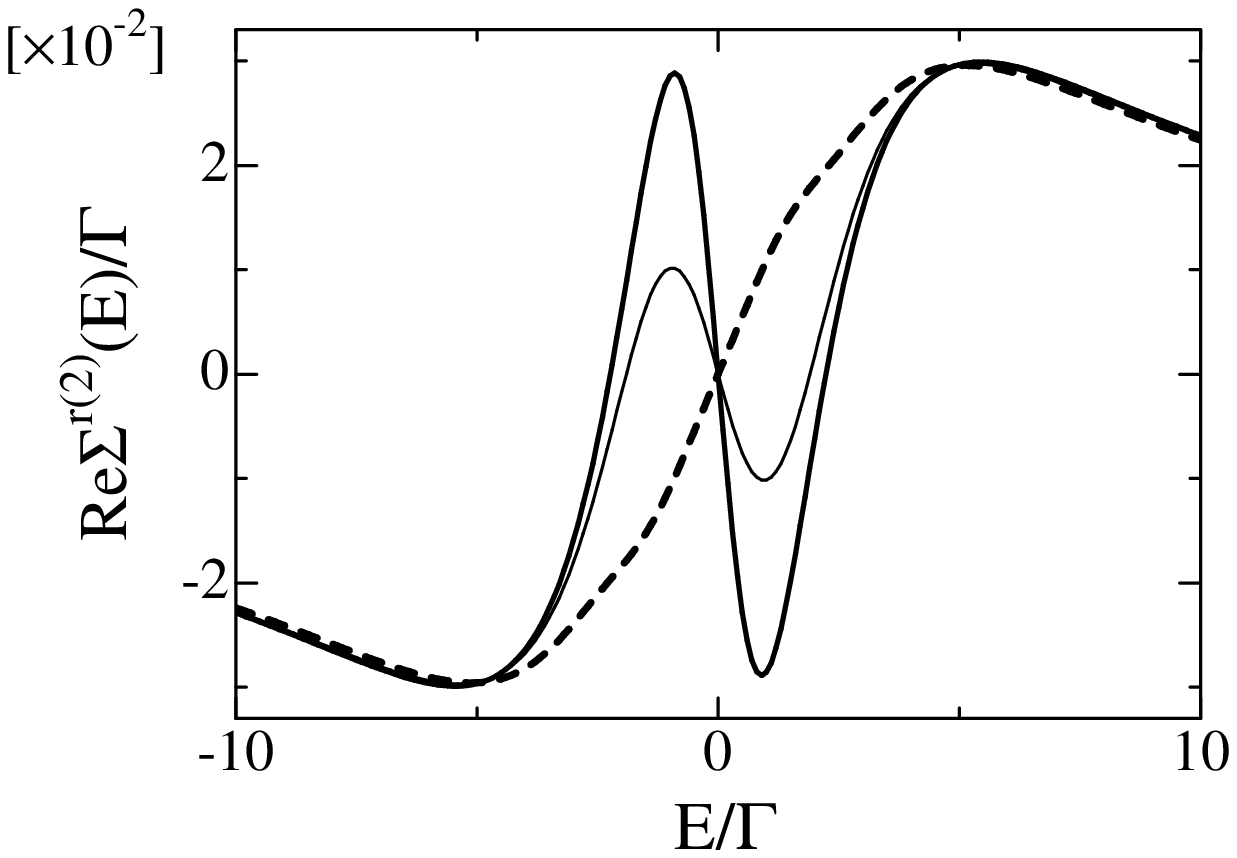}
\includegraphics[width=8.0cm]{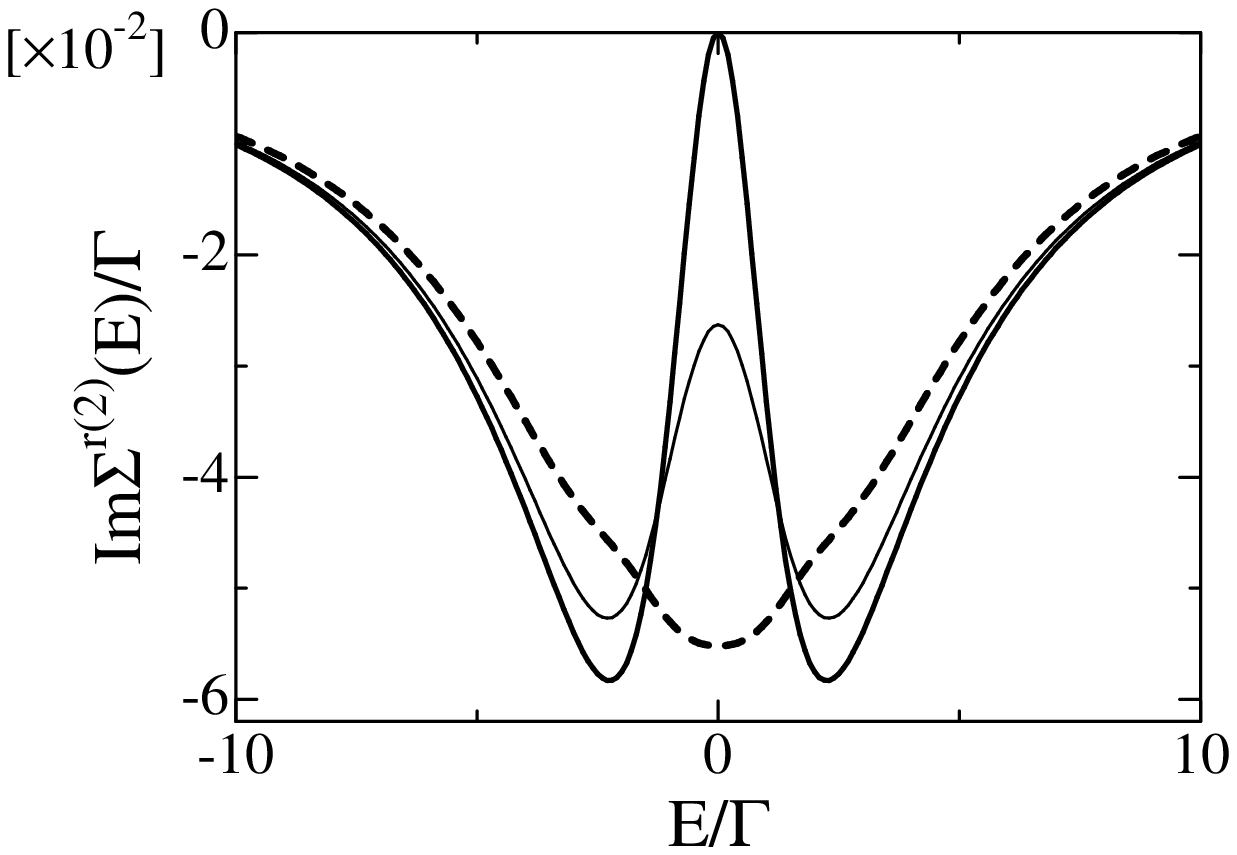}
\caption{
The second-order self-energy for the symmetric Anderson model 
at $U/{\Gamma}=1.0$ and zero temperature. 
(a) The real part and (b) the imaginary part  
 at equilibrium ( solid line ), 
$eV/{\Gamma}=1.0$ ( thin solid line ), and 
$eV/{\Gamma}=2.0$ ( dashed line ). 
 }
\includegraphics[width=8.0cm]{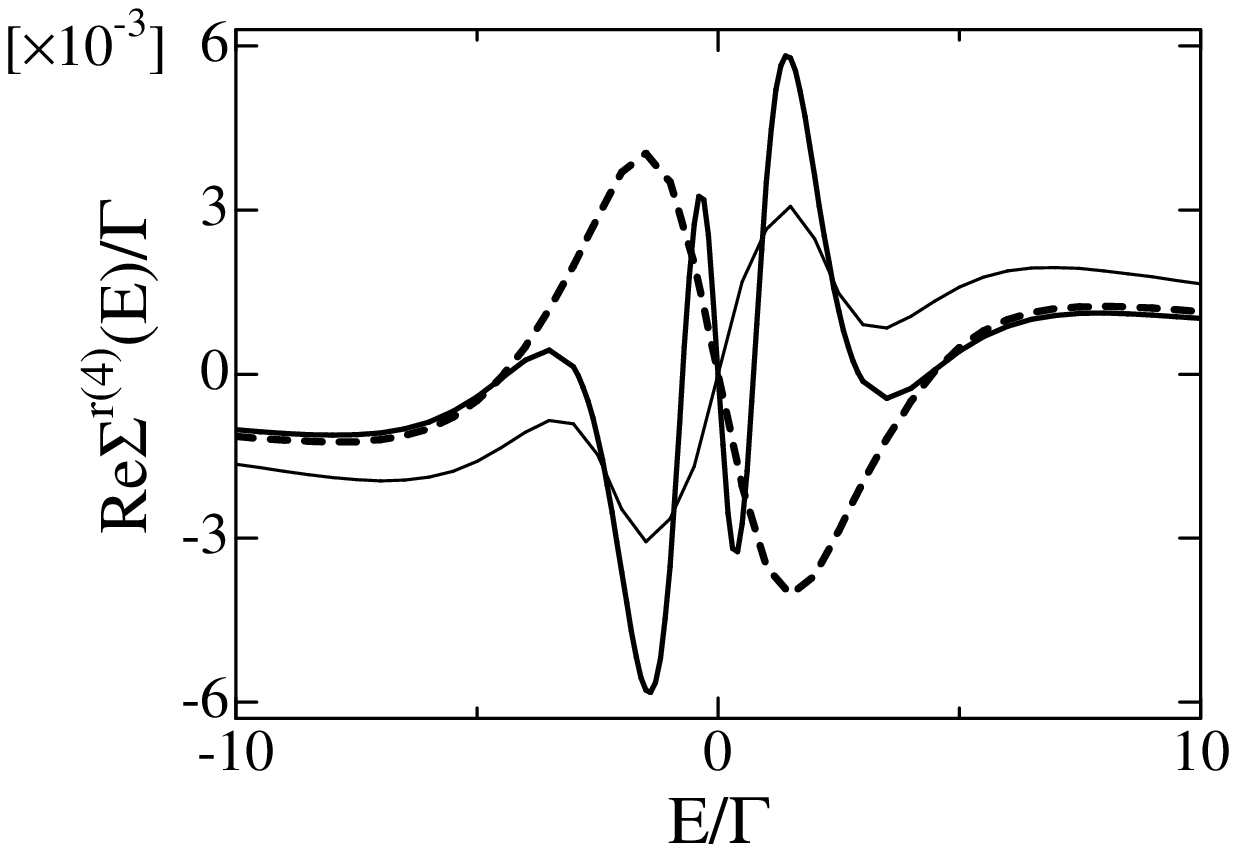}
\includegraphics[width=8.0cm]{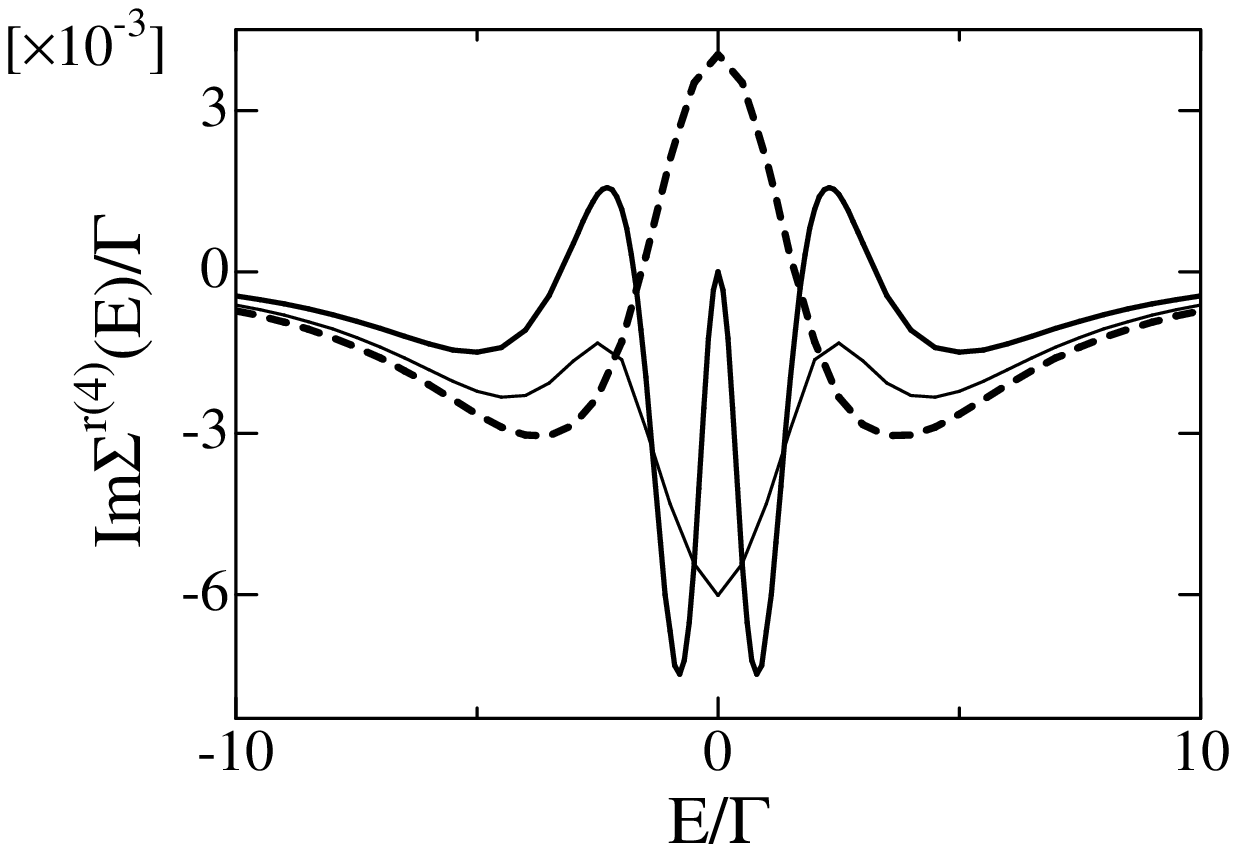}
\caption{
The fourth-order self-energy for the symmetric Anderson model 
at $U/{\Gamma}=1.0$ and zero temperature. 
(a) The real part and (b) the imaginary part 
at equilibrium ( solid line ), 
$eV/{\Gamma}=1.0$ ( thin solid line ), and 
$eV/{\Gamma}=2.0$ ( dashed line ). 
The fourth-order contribution for equilibrium 
has the same but narrow curves at low energy with those of the second-order 
contribution. }
\end{figure}

 The second-order and the fourth-order contributions to self-energy 
for zero temperature symmetric Anderson model 
are shown in Figs. 5(a) and 5(b) 
and in Figs. 6(a) and 6(b), respectively. 
Equation ( 27 ) represents  the curves around $E=0$ 
  denoted by solid line in Figs. 5(a) and 5(b), respectively, 
and Equation ( 32 ) represents approximately 
those  shown in Figs. 6(a) and 6(b), 
respectively. The curves of the second-order self-energy shown  
 in Figs. 5(a) and 5(b) are identical with those of expressions 
derived by Hershfield {\it et al.}~\cite{15}. 
In comparison of Figs. 6(a) and 6(b) with Figs. 
5(a) and 5(b), it is found that the fourth-order contribution for equilibrium 
has the same but narrow curves at low energy with those of the second-order 
contribution.  In addition, the broad curves are attached 
at high energy for the fourth-order self-energy. 
( The higher-order contribution is, the more the curves 
must oscillate as a function of energy. ) 
When the voltage, $eV/{\Gamma}$ exceeds ${\sim}2.0$, 
the behavior of curves of self-energy changes distinctly and  
 comes to present striking contrasts to that for the second-order 
contribution. Especially, the curve for the imaginary part 
of the fourth-order contribution  rises up  
with maximum at $E=0$.   
On the other hand, for the second-order contribution, 
 a valley appears with minimum at energy of zero$-$it  
 is quite the contrary. Moreover, 
from these results, 
it is expected that the sixth-order contribution 
to imaginary part of self-energy has minimum at $E=0$. 
Because of these, the perturbative expansion is 
hard to converge for $eV/{\Gamma}>{\sim}2.0$, as 
mentioned later.

Besides, the current conservation is mentioned. 
In Ref. [15], it is shown that the continuity of current 
entering and leaving the impurity  
 stands exactly at any strength of $U$   
within the approximation up to the second-order 
for the symmetric single-impurity Anderson model.  
In comparison of Figs. 6(a) and 6(b) with Figs. 
5(a) and 5(b), it is found that curves of fourth-order 
self-energy have the symmetry similar to those of 
the second-order. 
From this, it is anticipated that the current conservation are 
satisfied perfectly with approximation up to the 
fourth-order in the single-impurity system where 
electron-hole symmetry holds. 
The continuity of current can be maintained   
perfectly in single-impurity system as far as electron-hole 
symmetry stands. On the other hand, current comes to fail 
to be conserved with increasing $U$ 
in asymmetric single-impurity case and 
in two-impurity case. 

\subsection{ Spectral Function }

The spectral function with the second-order self-energy 
is generally known. 
It is  plotted for $U/{\Gamma}=10.0$ and zero temperature in Fig. 7.  
For equilibrium, the Kondo peak at energy of zero is very sharp 
and the two-side broad peaks appear at $E {\simeq}$ ${\pm}U/2$. 
The curve for $eV=0$ is identical with that shown in Ref. [24]. 
As $eV$ becomes higher than the Kondo temperatures, 
 $k_BT_K$~\cite{27}, 
the Kondo peak becomes lower and finally vanishes, 
while the two-side broad peaks rise at $E{\simeq}$ 
${\pm}U/2$.~\cite{15} 
\begin{figure}[htbp]
\begin{center}
\includegraphics[width=8.0cm]{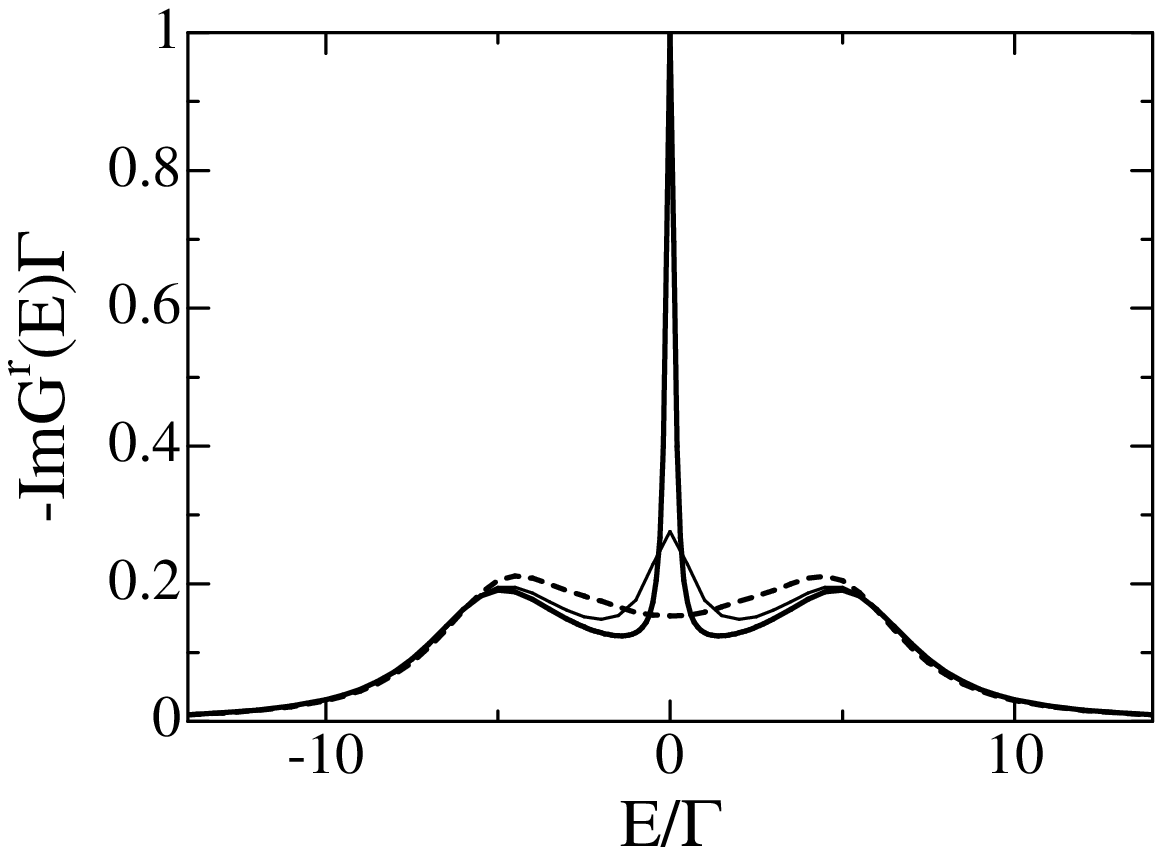}
\caption{
The spectral function with the second-order self-energy  
at  $U/{\Gamma}=10.0$ for the symmetric Anderson model  
 at equilibrium ( solid line ), 
$eV/{\Gamma}=1.0$ ( thin solid line ) and 
$eV/{\Gamma}=2.0$ ( dashed line ).   }
\includegraphics[width=8.0cm]{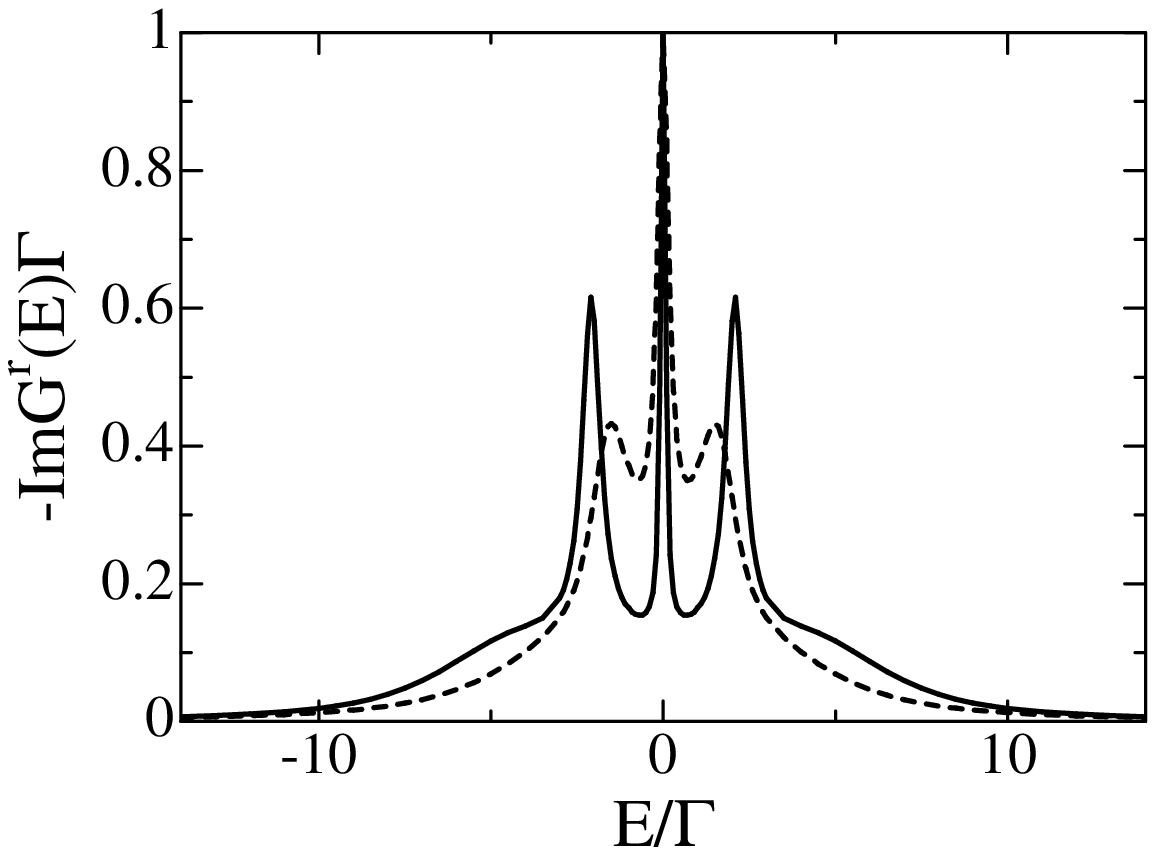}
\caption{
The spectral function with  self-energy  up to the fourth-order
at equilibrium for the symmetric Anderson model 
 at $U/{\Gamma}=3.5$ ( dashed line ) and  
$U/{\Gamma}=5.0$ ( solid line ).     }
\end{center}
\end{figure}

Figure 8 shows the spectral function with the self-energy  
 up to the fourth-order for equilibrium and zero temperature.
With strengthening $U$, two-side narrow peaks come to occur 
 in the vicinity of $E=$ ${\pm}U/2$ in addition to the Kondo peak. 
At $U$ large enough, the Kondo peak becomes very acute 
and two-side narrow peaks rise higher and sharpen; 
the energy levels for the atomic limit are  produced distinctly.  
The fourth-order self-energy has the same but narrow curves 
 as functions of energy with those of the second-order and 
those curves make the peaks at $E=$ ${\pm}U/2$. 

For the present approximation up to the fourth-order, 
the Kondo peak at $E=0$ reaches 
the unitarity limit and the charge, ${\langle}n{\rangle}$ 
corresponds to $1/2$, that is, 
the Friedel sum rule is correctly satisfied:~\cite{28}  
\begin{eqnarray}
{\rho}(E_f)={\sin}^2({\pi}{\langle}n{\rangle})/
{\pi}{\Gamma}, 
\end{eqnarray}
where ${\rho}(E_f)$ is the local density of states at 
the Fermi energy. Here, 
the discussions should be made on the ranges of $U$ in which the 
present approximation up to the fourth-order stands. 
From the results,  it is found that the approximation within  
the  fourth-order holds up to $U/{\Gamma}$
${\sim}5.0$ and is beyond the validity for $U/{\Gamma}$$>{\sim}6.0$.  
In addition, the curve for imaginary part of the fourth-order 
contribution is positive partly, as shown in Fig. 6(b) and 
as a consequence, the curve of the spectral function 
becomes negative partly for  too large $U$. In such a case, the present 
approximation is out of validity and the higher-order terms are required.  

\begin{figure}[htbp]
\includegraphics[width=8.0cm]{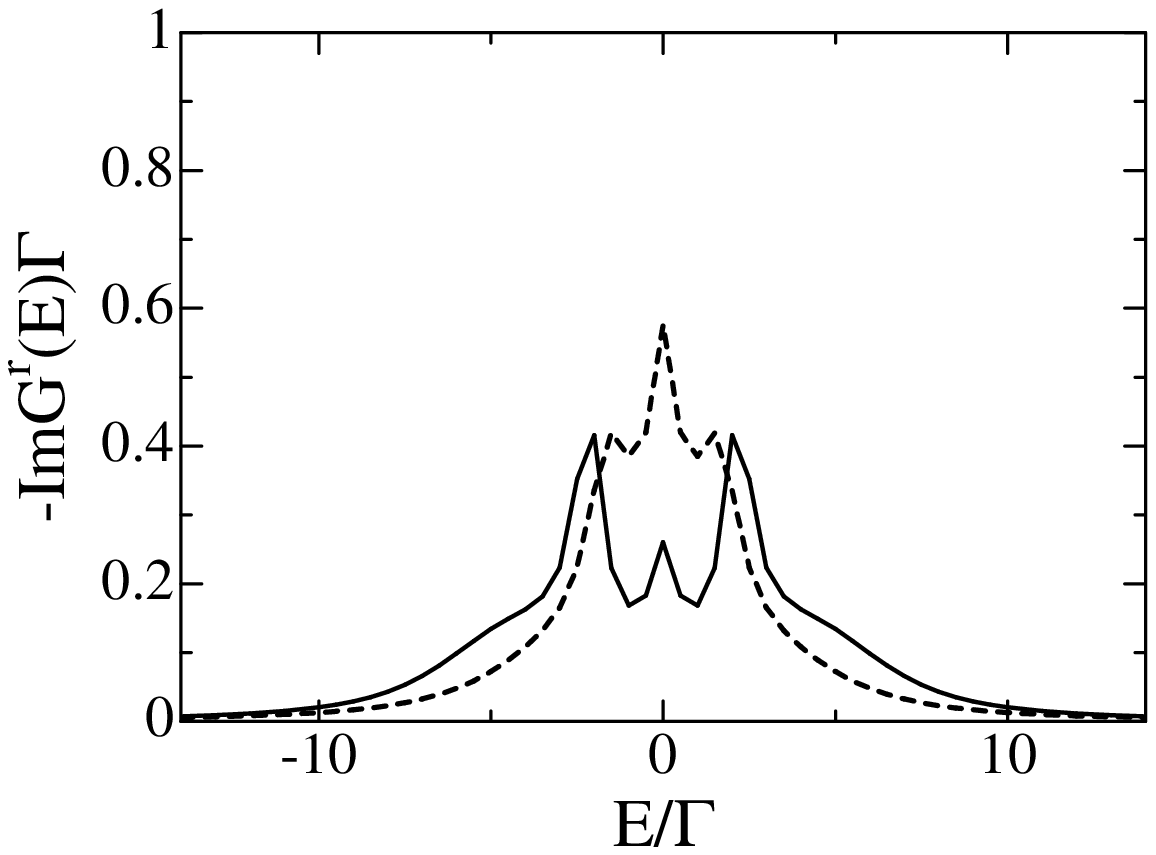}%
\includegraphics[width=8.0cm]{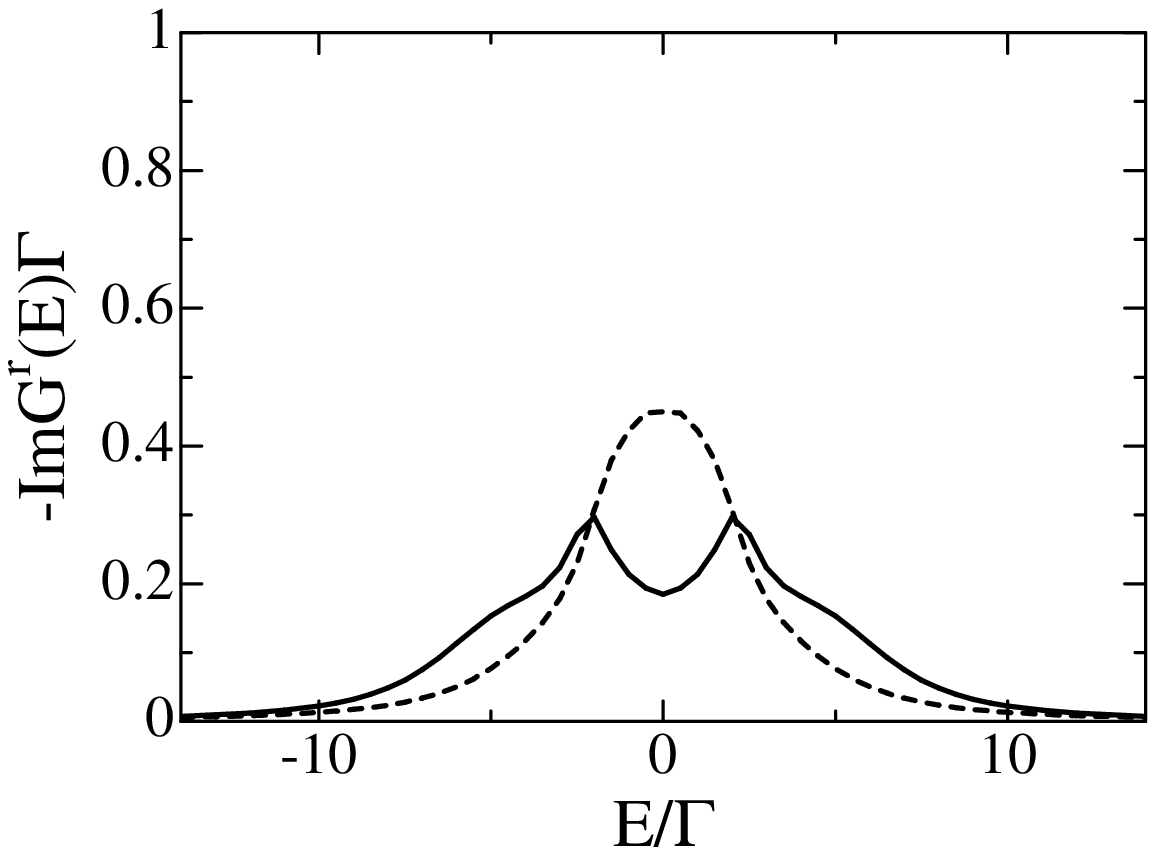}%
\caption{The spectral functions with  self-energy  up to the fourth-order  
at $eV/{\Gamma}=0.5$ ( Left ) and $eV/{\Gamma}=1.0$ ( Right )
for the symmetric Anderson model.  
$U/{\Gamma}=3.5$ ( dashed line ) and 
$U/{\Gamma}=5.0$ (  solid line ).   }
\end{figure}
Next, the results for nonequilibrium and zero temperature are 
shown. The expression for the Friedel sum rule, Eq. ( 33 ) 
does not stand for nonequilibrium,  since the charge cannot be 
expressed with respect to  the local density of states.
 All the same, the Kondo peak reaches the unitarity limit 
and ${\langle}n{\rangle}=1/2$ in the symmetric and noninteracting case.
The spectral functions with the self-energy up to the fourth-order 
 are plotted for $eV/{\Gamma}=0.5$ and $eV/{\Gamma}=1.0$ 
in Figs. 9, respectively. 
When $U$ is strengthened and $eV$ exceeds $k_BT_K$ 
( approximately, $k_BT_K/$${\Gamma}$ ${\sim}$$0.5$ for $U/$${\Gamma}$
$=3.5$ and $k_BT_K/$${\Gamma}$ ${\sim}$$0.3$ for $U/$${\Gamma}$$=5.0$ ),  
the Kondo peak for $eV/{\Gamma}=0.5$ falls in and   
instead, the two-side narrow peaks remain to sharpen in the vicinity 
of $E=$ ${\pm}U/2$. For $eV/{\Gamma}=1.0$, 
the Kondo peak becomes broad and disappears for $U$ large enough. 
The two-side peaks is generated small in the vicinity of $E=$ ${\pm}U/2$. 
The Kondo resonance is quite broken for bias voltage exceeding the Kondo temperatures;  
this accords with the experimental results of two terminal systems  
that the Kondo effect is suppressed when source-drain bias voltage is 
comparable to or exceeds the Kondo temperatures,  
 $eV  {\ge} k_BT_K$.~\cite{16,17}
For $eV/{\Gamma}>{\sim}2.0$, the Kondo peak does not lower  
even when $eV$ is much larger than $k_BT_K$. 
The perturbative expansion is hard to converge 
on account of the imaginary part of the self-energy 
for $eV/{\Gamma}>{\sim}2.0$, as described before; thereby, 
the higher-order contribution to self-energy is probably 
needed for high voltage. 

In the present work, nonequilibrium state is represented 
as the superposition of the two leads and 
the effective Fermi distribution function, Eq. ( 24 ) 
is qualitatively similar to that for finite temperatures. 
From the analogy in the Fermi distribution function, 
it is inferred that 
there are nonequilibrium fluctuations similar to  
 thermal fluctuations.~\cite{29} Because of the effective Fermi 
distribution function, not only for the second-order 
but also for the fourth-order, the Kondo resonance is destroyed, 
qualitatively the same as for finite temperatures. 
In contrast, if the finite voltage state 
is expressed as two localized states,     
the numerical results of the Kondo peak splitting 
can be obtained.   
All the same, for two terminal systems, the Kondo resonance 
splitting may not take place for finite bias voltage.  

{\it Summary}: The present work is based on 
the nonequilibrium perturbative formalism. 
Here the self-energies are derived  
and then it is indicated that the nonequilibrium ( real-time ) 
perturbative expansion  can be related to the Matsubara imaginary-time 
perturbative expansion for equilibrium. As the numerical results, 
the Kondo peak disappears as bias voltage exceeding the Kondo temperatures.  
Because of the analogy of the effective Fermi distribution function 
for nonequilibrium with that for finite temperatures, 
the present result is qualitatively similar to that for finite temperatures.  
This characteristic appears in the experiments of two terminal systems.

 \ \ 
  
\section*{Acknowledgements}
The numerical calculations were executed 
at the Yukawa Institute Computer Facility. 
The multiple integrals were performed   
using the  computer subroutine, {\it MQFSRD} of NUMPAC.  
 
\section*{Appendix }

The twelve terms for the fourth-order contribution  
 can be divided into four groups, each of which is composed  
of three terms. The four groups are brought from diagrams 
 denoted in  Figs. 4 (a)-(c),  Figs. 4 (d)-(f), Figs. 4 (g)-(i), 
and Figs. 4  (j)-(l), respectively. 
The terms for the diagrams illustrated 
in Figs. 4(a) and 4(b) are equivalent 
except for the spin indices and expressed by 
\begin{eqnarray}
 {\Sigma}^{r(4)}_{a, b}(E)
&=&U^4\int^{\infty}_{0} {dt_{1}}\int^{\infty}_{-\infty}
{dt_{2}}\int^{\infty}_{-\infty}{dt_{3}}\, e^{iEt_1} \nonumber \\ 
  \ \  & & {\times} \left[ 
\begin{array}
{llll}
 g^<(t_1)
 g^<(t_1-t_2-t_3)
 g^>(-t_1+t_2+t_3) \nonumber \\ 
 -  g^>(t_1)
 g^>(t_1-t_2-t_3)
 g^<(-t_1+t_2+t_3) \nonumber \\ 
\end{array} 
\right] \nonumber \\
  \ \  & & {\times}  
\left[ \begin{array}
{llll}
 g^{\pm}(t_2)
 g^<(-t_2)  
 + g^<(t_2)
 g^{\pm}(-t_2) 
\end{array} 
\right] \nonumber \\
  \ \  & & {\times}\left[ \begin{array}
{llll}
 g^{\pm}(t_3)
 g^<(-t_3)  
 + g^<(t_3)
 g^{\pm}(-t_3)  
\end{array} 
\right],  
\end{eqnarray}
\begin{eqnarray}
 {\Sigma}^{a(4)}_{a, b}(E)
&=&U^4\int^{0}_{-\infty} {dt_{1}}\int^{\infty}_{-\infty}
{dt_{2}}\int^{\infty}_{-\infty}{dt_{3}}\, e^{iEt_1}
\nonumber \\   \ \  & & {\times}\left[ 
\begin{array}
{llll}
 g^>(t_1)
 g^>(t_1-t_2-t_3)
 g^<(-t_1+t_2+t_3) \nonumber \\ 
-g^<(t_1)
 g^<(t_1-t_2-t_3)
 g^>(-t_1+t_2+t_3) \nonumber \\ 
\end{array} 
\right] \\
 \ \     & &{\times} 
\left[ \begin{array}
{llll}
 g^{\pm}(t_2)
 g^<(-t_2)  
 + g^<(t_2)
 g^{\pm}(-t_2) 
\end{array} 
\right]   \nonumber \\ 
\ \    & &{\times}\left[ \begin{array}
{llll}
 g^{\pm}(t_3)
 g^<(-t_3)  
 + g^<(t_3)
 g^{\pm}(-t_3)  
\end{array} 
\right]. 
\end{eqnarray}
Additionally, Figure  4(c) shows the diagram 
for the following terms:  
\begin{eqnarray}
 {\Sigma}^{r(4)}_{c}(E)
&=&U^4\int^{\infty}_{0} {dt_{1}}\int^{\infty}_{-\infty}
{dt_{2}}\int^{\infty}_{-\infty}{dt_{3}}\, e^{iEt_1}
\nonumber \\ & & {\times} \left[ 
\begin{array}
{llll}
 g^>(-t_1)
 g^<(t_1-t_2-t_3)
 g^<(t_1-t_2-t_3) \nonumber \\ 
 -  g^<(-t_1)
 g^>(t_1-t_2-t_3)
 g^>(t_1-t_2-t_3) \nonumber \\ 
\end{array} 
\right] \\
  \ \  & & {\times} 
\left[ \begin{array}
{llll}
 g^{\pm}(t_2)
 g^>(t_2)  
 + g^<(t_2)
 g^{\pm}(t_2)  \end{array} 
\right]  \nonumber \\
  \ \  & & {\times}
\left[ \begin{array}
{llll}
 g^{\pm}(t_3)
 g^>(t_3)  
 + g^<(t_3)
 g^{\pm}(t_3)  \end{array} 
\right], 
\end{eqnarray}
\begin{eqnarray}
 {\Sigma}^{a(4)}_{c}(E)
&=&U^4\int^{0}_{-\infty} {dt_{1}}\int^{\infty}_{-\infty}
{dt_{2}}\int^{\infty}_{-\infty}{dt_{3}}\, e^{iEt_1}
\nonumber \\  & & {\times} \left[ 
\begin{array}
{llll}
 g^<(-t_1)
 g^>(t_1-t_2-t_3)
 g^>(t_1-t_2-t_3) \nonumber \\ 
 -g^>(-t_1)
 g^<(t_1-t_2-t_3)
 g^<(t_1-t_2-t_3) \nonumber \\ 
\end{array} 
\right] \\
 \ \   & & {\times} 
\left[ \begin{array}
{llll}
 g^{\pm}(t_2)
 g^>(t_2)  
 + g^<(t_2)
 g^{\pm}(t_2)  \end{array} 
\right] \nonumber \\
 \ \  & & {\times} 
\left[ \begin{array}
{llll}
 g^{\pm}(t_3)
 g^>(t_3)  
 + g^<(t_3)
 g^{\pm}(t_3)  \end{array} 
\right]. 
\end{eqnarray}
Next, the terms brought from diagram  
 in Fig. 4(d) are expressed by 
\begin{eqnarray}
 {\Sigma}^{r(4)}_{d}(E)
&=&U^4\int^{\infty}_{0} {dt_{1}}\int^{\infty}_{-\infty}
{dt_{2}}\int^{\infty}_{-\infty}{dt_{3}}\, e^{iEt_1}
\nonumber \\   & & \ \  {\times} \left[ 
\begin{array}
{llll}
 g^>(t_1-t_3)
 g^>(t_1-t_2) g^<(t_2-t_1)
 \nonumber \\ 
 -  g^<(t_1-t_3) g^<(t_1-t_2)
 g^>(t_2-t_1)
 \nonumber \\ 
\end{array} 
\right] \\
 & & \ \  {\times}
 \,g^{\pm}(t_2)\,{\rm sgn}(t_3)
\left[ \begin{array}
{llll}
 g^>(-t_2+t_3) g^>(t_3)
 g^<(-t_3)   \\ 
 -   g^<(-t_2+t_3)g^<(t_3)
 g^>(-t_3)
\end{array}
\right], 
\end{eqnarray}
\begin{eqnarray}
 {\Sigma}^{a(4)}_{d}(E)
&=&U^4\int^{0}_{-\infty} {dt_{1}}\int^{\infty}_{-\infty}
{dt_{2}}\int^{\infty}_{-\infty}{dt_{3}}\, e^{iEt_1}
\nonumber \\ & & {\times} \left[ 
\begin{array}
{llll}
g^<(t_1-t_3) g^<(t_1-t_2)
 g^>(t_2-t_1)
 \nonumber \\ 
-g^>(t_1-t_3)
 g^>(t_1-t_2) g^<(t_2-t_1)
 \nonumber \\ 
\end{array} 
\right] \\
& & {\times} \,g^{\pm}(t_2)\,{\rm sgn}(t_3)
\left[ 
\begin{array}
{llll}
 g^>(-t_2+t_3) g^>(t_3)
 g^<(-t_3)    \\ 
 -   g^<(-t_2+t_3)g^<(t_3)
 g^>(-t_3)
\end{array}
\right]. 
\end{eqnarray}
The terms for diagram in Fig. 4(e) are written by 
\begin{eqnarray}
 {\Sigma}^{r(4)}_{e}(E)
&=&U^4\int^{\infty}_{0} {dt_{1}}\int^{\infty}_{-\infty}
{dt_{2}}\int^{\infty}_{-\infty}{dt_{3}}\, e^{iEt_1}
\nonumber \\ & &   {\times}\left[ 
\begin{array}
{llll}
 g^>(t_1-t_2)
 g^>(t_1-t_2) g^<(t_3-t_1)
 \nonumber \\ 
 -  g^<(t_1-t_2) g^<(t_1-t_2)
 g^>(t_3-t_1)
 \nonumber \\ 
\end{array} 
\right] \\
 & &  {\times} \,g^{\pm}(t_2)\,{\rm sgn}(t_3)
\left[ 
\begin{array}
{llll}
 g^>(t_2-t_3) g^>(-t_3)
 g^<(t_3)    \\ 
 -   g^<(t_2-t_3)g^<(-t_3)
 g^>(t_3) \\
\end{array}
\right], 
\end{eqnarray}
\begin{eqnarray}
 {\Sigma}^{a(4)}_{e}(E)
&=&U^4\int^{0}_{-\infty} {dt_{1}}\int^{\infty}_{-\infty}
{dt_{2}}\int^{\infty}_{-\infty}{dt_{3}}\, e^{iEt_1} 
\nonumber \\ & & {\times} \left[ 
\begin{array}
{llll}
 g^<(t_1-t_2) g^<(t_1-t_2)
 g^>(t_3-t_1)
 \nonumber \\ 
- g^>(t_1-t_2)
 g^>(t_1-t_2) g^<(t_3-t_1)
 \nonumber \\ 
\end{array} 
\right] \\
  & & {\times} \,g^{\pm}(t_2)\,{\rm sgn}(t_3)
\left[ 
\begin{array}
{llll}
 g^>(t_2-t_3) g^>(-t_3)
 g^<(t_3)    \\ 
 -   g^<(t_2-t_3)g^<(-t_3)
 g^>(t_3)\\
\end{array}
\right].  
\end{eqnarray}
In addition, Figure 4(f) denotes the diagram for the 
following terms: 
\begin{eqnarray}
 {\Sigma}^{r(4)}_{f}(E)
&=&U^4\int^{\infty}_{0} {dt_{1}}\int^{\infty}_{-\infty}
{dt_{2}}\int^{\infty}_{-\infty}{dt_{3}}\, e^{iEt_1}
\nonumber \\ & & {\times} \left[ 
\begin{array}
{llll}
 g^>(t_1-t_3)
 g^>(t_1-t_2) g^<(t_2-t_1)
 \nonumber \\ 
 -  g^<(t_1-t_3) g^<(t_1-t_2)
 g^>(t_2-t_1)
 \nonumber \\ 
\end{array} 
\right] \\
    & &{\times} \,g^{\pm}(-t_2)\,{\rm sgn}(t_3)
\left[ 
\begin{array}
{llll}
 g^<(t_3)g^<(t_3)
 g^>(t_2-t_3)   \\ 
- g^>(t_3) g^>(t_3)
 g^<(t_2-t_3)  
\end{array}
\right], 
\end{eqnarray}
\begin{eqnarray}
 {\Sigma}^{a(4)}_{f}(E)
&=&U^4\int^{0}_{-\infty} {dt_{1}}\int^{\infty}_{-\infty}
{dt_{2}}\int^{\infty}_{-\infty}{dt_{3}}\, e^{iEt_1} 
\nonumber \\
& & {\times}
\left[ 
\begin{array}
{llll}
 g^<(t_1-t_3) g^<(t_1-t_2)
 g^>(t_2-t_1)
 \nonumber \\ 
- g^>(t_1-t_3)
 g^>(t_1-t_2) g^<(t_2-t_1)
 \nonumber \\ 
\end{array} 
\right] \\
   & &{\times} \,g^{\pm}(-t_2)\,{\rm sgn}(t_3)
\left[ 
\begin{array}
{llll}
 g^<(t_3)g^<(t_3)
 g^>(t_2-t_3)   \\ 
  - g^>(t_3) g^>(t_3)
 g^<(t_2-t_3)  
\end{array}
\right]. 
\end{eqnarray}
Next, the terms formulated from diagram   
illustrated in Fig. 4(g) are expressed by 
\begin{eqnarray}
 {\Sigma}^{r(4)}_{g}(E)
&=&U^4\int^{\infty}_{0} {dt_{1}}\int^{\infty}_{-\infty}
{dt_{2}}\int^{\infty}_{-\infty}{dt_{3}}\, e^{iEt_1}
\nonumber \\
& & {\times}
\left[ 
\begin{array}
{llll}
 g^>(t_1)
 g^>(t_1-t_2-t_3)
 g^<(t_2-t_1) \nonumber \\ 
 -  g^<(t_1)
 g^<(t_1-t_2-t_3)
 g^>(t_2-t_1) \nonumber \\ 
\end{array} 
\right] \\
 \ \   & &{\times} \,g^{\pm}(-t_2)\,{\rm sgn}(t_3)
\left[ 
\begin{array}
{llll}
  g^>(t_2+t_3)g^>(t_3)
 g^<(-t_3)   \\ 
 -   g^<(t_2+t_3)g^<(t_3)
 g^>(-t_3)
\end{array}
\right], 
\end{eqnarray}
\begin{eqnarray}
 {\Sigma}^{a(4)}_{g}(E)
&=&U^4\int^{0}_{-\infty} {dt_{1}}\int^{\infty}_{-\infty}
{dt_{2}}\int^{\infty}_{-\infty}{dt_{3}}\, e^{iEt_1}
\nonumber \\
& & {\times}
\left[ 
\begin{array}
{llll}
 g^<(t_1)
 g^<(t_1-t_2-t_3)
 g^>(t_2-t_1) \nonumber \\ 
- g^>(t_1)
 g^>(t_1-t_2-t_3)
 g^<(t_2-t_1) \nonumber \\ 
\end{array} 
\right] \\
 \ \   & & {\times} \,g^{\pm}(-t_2)\,{\rm sgn}(t_3)
\left[ 
\begin{array}
{llll}
 g^>(t_2+t_3) g^>(t_3)
 g^<(-t_3)   \\ 
 -  g^<(t_2+t_3)g^<(t_3)
 g^>(-t_3)
\end{array}
\right]. 
\end{eqnarray}
Figure 4(h) illustrates the diagram for 
the following terms:
\begin{eqnarray}
 {\Sigma}^{r(4)}_{h}(E)
&=&U^4\int^{\infty}_{0} {dt_{1}}\int^{\infty}_{-\infty}
{dt_{2}}\int^{\infty}_{-\infty}{dt_{3}}\, e^{iEt_1}
\nonumber \\
& & {\times}
\left[ 
\begin{array}
{llll}
 g^<(t_1)
 g^<(t_1-t_2-t_3)
 g^>(t_2-t_1) \nonumber \\ 
 -  g^>(t_1)
 g^>(t_1-t_2-t_3)
 g^<(t_2-t_1) \nonumber \\ 
\end{array} 
\right] \\
 \ \   & & {\times} \,g^{\pm}(t_2)\,{\rm sgn}(t_3)
\left[ 
\begin{array}
{llll}
 g^>(t_3)
 g^>(t_3)
 g^<(-t_2-t_3)    \\ 
 -  g^<(t_3)
 g^<(t_3)
 g^>(-t_2-t_3)  
\end{array}
\right], 
\end{eqnarray}
\begin{eqnarray}
 {\Sigma}^{a(4)}_{h}(E)
&=&U^4\int^{0}_{-\infty} {dt_{1}}\int^{\infty}_{-\infty}
{dt_{2}}\int^{\infty}_{-\infty}{dt_{3}}\, e^{iEt_1}
\nonumber \\
& & {\times}
\left[ 
\begin{array}
{llll}
 g^>(t_1)
 g^>(t_1-t_2-t_3)
 g^<(t_2-t_1) \nonumber \\ 
-g^<(t_1)
 g^<(t_1-t_2-t_3)
 g^>(t_2-t_1) \nonumber \\ 
\end{array} 
\right] \\
 \ \   & & {\times} \,g^{\pm}(t_2)\,{\rm sgn}(t_3)
\left[ 
\begin{array}
{llll}
 g^>(t_3)
 g^>(t_3)
 g^<(-t_2-t_3)    \\ 
 -  g^<(t_3)
 g^<(t_3)
 g^>(-t_2-t_3)  
\end{array}
\right]. 
\end{eqnarray}
Besides, the terms formulated from the diagram in Fig. 4(i) 
 are written by 
\begin{eqnarray}
 {\Sigma}^{r(4)}_{i}(E)
&=&U^4\int^{\infty}_{0} {dt_{1}}\int^{\infty}_{-\infty}
{dt_{2}}\int^{\infty}_{-\infty}{dt_{3}}\, e^{iEt_1}
\nonumber \\
& & {\times}
\left[ 
\begin{array}
{llll}
 g^>(-t_1)
 g^<(t_1-t_2-t_3)
 g^<(t_1-t_2) \nonumber \\ 
 -  g^<(-t_1)
 g^>(t_1-t_2-t_3)
 g^>(t_1-t_2) \nonumber \\ 
\end{array} 
\right] \\
 \ \   & & {\times} \,g^{\pm}(t_2)\,{\rm sgn}(t_3)
\left[ 
\begin{array}
{llll}
  g^>(t_2+t_3)g^>(t_3)
 g^<(-t_3)   \\ 
 - g^<(t_2+t_3) g^<(t_3)
 g^>(-t_3)
\end{array}
\right], 
\end{eqnarray}
\begin{eqnarray}
 {\Sigma}^{a(4)}_{i}(E)
&=&U^4\int^{0}_{-\infty} {dt_{1}}\int^{\infty}_{-\infty}
{dt_{2}}\int^{\infty}_{-\infty}{dt_{3}}\, e^{iEt_1}
\nonumber \\
& & {\times}
\left[ 
\begin{array}
{llll}
 g^<(-t_1)
 g^>(t_1-t_2-t_3)
 g^>(t_1-t_2) \nonumber \\ 
- g^>(-t_1)
 g^<(t_1-t_2-t_3)
 g^<(t_1-t_2) \nonumber \\ 
\end{array} 
\right] \\
 \ \   & & {\times} \,g^{\pm}(t_2)\,{\rm sgn}(t_3)
\left[ 
\begin{array}
{llll}
  g^>(t_2+t_3)g^>(t_3)
 g^<(-t_3)    \\ 
 -  g^<(t_2+t_3)  g^<(t_3)
 g^>(-t_3)
\end{array}
\right].
\end{eqnarray}
Next, the terms for diagrams denoted in Figs. 4 (j) and 4(k)  
are equivalent except for the spin indices and written by 
\begin{eqnarray}
 {\Sigma}^{r(4)}_{j, k}(E)
&=&U^4\int^{\infty}_{0} {dt_{1}}\int^{\infty}_{-\infty}
{dt_{2}}\int^{\infty}_{-\infty}{dt_{3}}\, e^{iEt_1}
\nonumber \\
& & {\times}
\left[ 
\begin{array}
{llll}
 g^>(t_1)
 g^<(-t_1)
 g^>(t_1-t_2-t_3) \nonumber \\ 
 -  g^<(t_1)
 g^>(-t_1)
 g^<(t_1-t_2-t_3) \nonumber \\ 
\end{array} 
\right] \\
 \ \   & & {\times} \,g^{\pm}(t_2)
\left[ 
\begin{array}
{llll}
\ \  g^{\pm}(t_3)
 g^>(t_3)
 g^<(-t_3)     \\
 +  g^<(t_3)
 g^{\pm}(t_3)
 g^>(-t_3)   \\
 +  g^<(t_3)
 g^>(t_3)
 g^{\pm}(-t_3)  \\ 
\end{array}
\right], 
\end{eqnarray}
\begin{eqnarray}
 {\Sigma}^{a(4)}_{j, k}(E)
&=&U^4\int^{0}_{-\infty} {dt_{1}}\int^{\infty}_{-\infty}
{dt_{2}}\int^{\infty}_{-\infty}{dt_{3}}\, e^{iEt_1}
\nonumber \\
& & {\times}
\left[ 
\begin{array}
{llll}
  g^<(t_1)
 g^>(-t_1)
 g^<(t_1-t_2-t_3) \nonumber \\ 
- g^>(t_1)
 g^<(-t_1)
 g^>(t_1-t_2-t_3) \nonumber \\ 
\end{array} 
\right] \\
 \ \   & & {\times} \,g^{\pm}(t_2)
\left[ 
\begin{array}
{llll}
\ \  g^{\pm}(t_3)
 g^>(t_3)
 g^<(-t_3)   \\
 +  g^<(t_3)
 g^{\pm}(t_3)
 g^>(-t_3)  \\ 
 +  g^<(t_3)
 g^>(t_3)
 g^{\pm}(-t_3)  \\ 
\end{array}
\right]. 
\end{eqnarray}
In addition, the terms for diagram illustrated in Fig. 4(l)  
are expressed by 
\begin{eqnarray}
 {\Sigma}^{r(4)}_{l}(E)
&=&U^4\int^{\infty}_{0} {dt_{1}}\int^{\infty}_{-\infty}
{dt_{2}}\int^{\infty}_{-\infty}{dt_{3}}\, e^{iEt_1}
\nonumber \\
& & {\times}
\left[ 
\begin{array}
{llll}
 g^>(t_1)
 g^>(t_1)
 g^<(-t_1+t_2+t_3) \nonumber \\ 
 -  g^<(t_1)
 g^<(t_1)
 g^>(-t_1+t_2+t_3) \nonumber \\ 
\end{array} 
\right] \nonumber \\
 \ \   & & {\times} \,g^{\pm}(-t_2)
\left[ 
\begin{array}
{llll}
\ \ g^{\pm}(-t_3)
 g^>(-t_3) g^<(t_3)  \\
 +  g^<(-t_3) g^{\pm}(-t_3)
 g^>(t_3) \\  +  g^<(-t_3)
 g^>(-t_3)
 g^{\pm}(t_3)   \end{array} 
\right],    
\end{eqnarray}
\begin{eqnarray}
 {\Sigma}^{a(4)}_{l}(E)
&=&U^4\int^{0}_{-\infty} {dt_{1}}\int^{\infty}_{-\infty}
{dt_{2}}\int^{\infty}_{-\infty}{dt_{3}}\, e^{iEt_1}
\nonumber \\
& & {\times}
\left[ 
\begin{array}
{llll}
 g^<(t_1)
 g^<(t_1)
 g^>(-t_1+t_2+t_3) \nonumber \\ 
-g^>(t_1)
 g^>(t_1)
 g^<(-t_1+t_2+t_3) \nonumber \\ 
\end{array} 
\right] \nonumber \\
 \ \   & & {\times} \,g^{\pm}(-t_2)
\left[ 
\begin{array}
{llll}
\ \ g^{\pm}(-t_3)
 g^>(-t_3) g^<(t_3) \\ 
 +  g^<(-t_3) g^{\pm}(-t_3)
 g^>(t_3) \\  +  g^<(-t_3)
 g^>(-t_3)
 g^{\pm}(t_3)   \end{array} 
\right]. 
\end{eqnarray}

\end{document}